\documentclass{article}

\usepackage{arxiv}
\usepackage[utf8]{inputenc} % allow utf-8 input
\usepackage[T1]{fontenc}    % use 8-bit T1 fonts
\usepackage{hyperref}       % hyperlinks
\usepackage{booktabs}       % professional-quality tables
\usepackage{amsfonts}       % blackboard math symbols
\usepackage{lipsum}
\usepackage{graphicx}
\usepackage{url}
\usepackage{stmaryrd}
\usepackage{amsmath}
\newcommand{\indep}{\perp \!\!\! \perp}
\usepackage{multicol}
\usepackage{multirow}

\title{Causal inference with multiple versions of treatment and application to personalized medicine}

\author{Jonas B\'eal\\
Institut Curie, PSL Research University, Inserm, U900\\
Paris, France\\
\texttt{jonas.beal@curie.fr}
\And
Aur\'elien Latouche\\
Institut Curie, Inserm U900\\
Conservatoire national des arts et m\'etiers\\
Paris, France
}

\begin{document}
\maketitle
\begin{abstract}
The development of high-throughput sequencing and targeted therapies has led to the emergence of personalized medicine: a patient's molecular profile or the presence of a specific biomarker of drug response will correspond to a treatment recommendation made either by a physician or by a treatment assignment algorithm. The growing number of such algorithms raises the question of how to quantify their clinical impact knowing that a personalized medicine strategy will inherently include different versions of treatment.

We thus specify an appropriate causal framework with multiple versions of treatment to define the causal effects of interest for precision medicine strategies and estimate them emulating clinical trials with observational data. Therefore, we determine whether the treatment assignment algorithm is more efficient than different control arms: gold standard treatment, observed treatments or random assignment of targeted treatments. 

Causal estimates of the precision medicine effects are first evaluated on simulated data and they demonstrate a lower biases and variances compared with naive estimation of the difference in expected outcome between treatment arms. The various simulations scenarios also point out the different bias sources depending on the clinical situation (heterogeneity of response, assignment of observed treatments etc.). A RShiny interactive application is also provided to further explore other user-defined scenarios. The method is then applied to data from patient-derived xenografts (PDX): each patient tumour is implanted in several immunodeficient cloned mice later treated with different drugs, thus providing access to all corresponding drug sensitivities for all patients. Access to these  unique pre-clinical data emulating counterfactual outcomes allows to validate the reliability of causal estimates obtained with the proposed method. 
\end{abstract}

% keywords can be removed

\section{Introduction}
\label{introduction}

Precision medicine (PM) consists in associating the most appropriate treatment to each patient according to his or her characteristics. This is, for instance, quite common in the clinical management of cancer patients where the choice of treatment is increasingly influenced by the genomic alterations of the patient \cite{friedman2015precision}. At the individual level, targeted treatments has provided relevant solutions for patients with specific mutations \cite{abou2003overview}. Putting together these various treatments, some precision medicine strategies can be defined: based on the omics profile of the patient, the treatment most likely to be successful is chosen. If the information available is reliable, precision medicine can thus be reduced to a treatment choice algorithm that takes as input the molecular characteristics of the patient's tumour and outputs a recommendation of treatment.

The question then arises of how to quantify the clinical benefit provided by these treatment algorithms. Some clinical trials have been proposed, demonstrating both the feasibility of collecting information about mutations \cite{le2015molecularly} or RNA \cite{rodon2019genomic} in real-time and the clinical benefit that can be expected from these approaches for some patients \cite{coyne2017defining}. However, the increasing abundance of omics data and biological knowledge make it progressively easier to establish new algorithms for precision medicine, either directly based on physician knowledge or provided by computational models \cite{hansen2013computation}. For practical reasons it is not possible to propose a real clinical trial for each new precision medicine algorithm or for any variants, comparing standard of care with new algorithm-based treatments. 

Therefore, this work provides a method to assess the clinical impact of proposed PM treatment algorithm based on already generated data, emulating clinical trials and analyzing them in the causal inference framework \cite{hernan2016using}. First we will define the causal estimates of the precision medicine effects (later referred to as causal estimates) we want to assess, and the corresponding ideal clinical trials one would like to perform. Next, we will define the notations and the causal framework we use to infer the causal effects from observational data with multiple versions of treatment, based on the previous work by \cite{vanderweele2013causal}. It will be briefly introduced in its main principles and then extended in order to adapt to the characteristics of PM, in particular the multiplicity of treatment versions, \emph{i.e.} targeted drugs. Then we will apply the proposed methods to simulated data in order to investigate the different biases of the candidate methods. An example scenario will be presented and a RShiny interactive application has been developed to further explore other user-defined settings. Finally, the analysis of data from patient-derived xenografts (PDX) makes it possible both to apply the methods to pre-clinical situation and to have data approximating the counterfactual responses, thus enabling further validation of the proposed estimation methods.

\section{Target trials for precision medicine: definition of causal estimates}\label{causal_estimates}

We first specify the precision medicine effects that are to be estimated. These effects will finally be estimated based on observational data through the causal framework and target trial emulation \cite{hernan2016using}. Thus, if we think in terms of clinical trials, we are not trying to prove or quantify the superiority of one treatment over another but rather to evaluate the clinical utility of a precision medicine strategy assigning treatments based on genomic features of patients. This is therefore closer to the well-studied biomarker-based designs for clinical trials \cite{freidlin2010randomized}. In a way, it is a matter of extending these unidimensional biomarker-based designs to multidimensional strategies that allow a choice between quite a number of different treatments. The potentially large number of treatments thus prompts us to draw more inspiration from scalable biomarker-strategy designs than biomarker-stratified designs \cite{freidlin2010randomized}. We can draw a methodological parallel with some trials like the Tumor Chemosensitivity Assay Ovarian Cancer study in which a biochemical assay guides the choice of preferred chemotherapy for patients in a panel of twelve different treatments \cite{cree2007prospective}. More recently, some clinical trials have been proposed that include precision medicine strategies, particularly in oncology \cite{le2015molecularly,flaherty2020molecular}. 

\begin{figure}[ht] 
\centering\includegraphics[width=\textwidth]{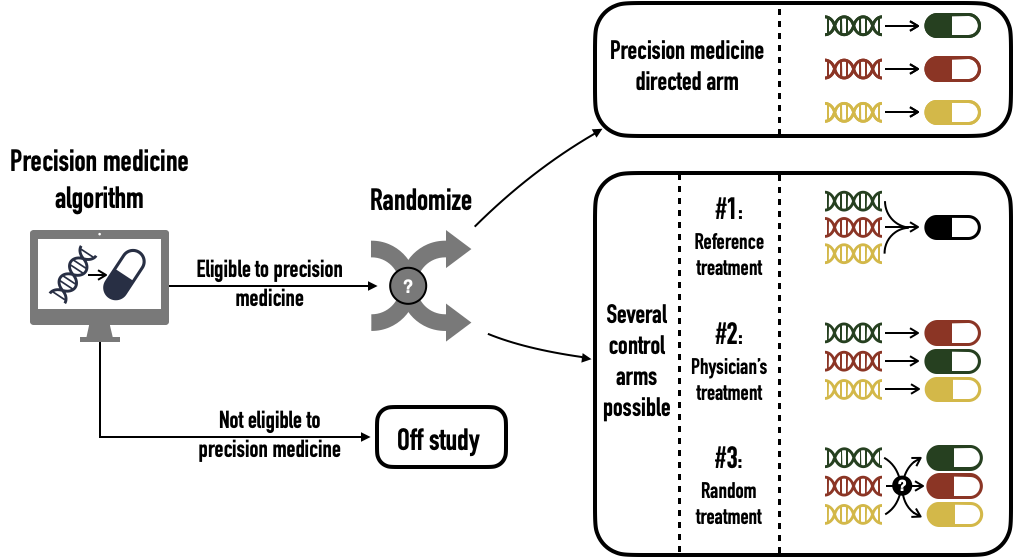}
\caption{\textbf{Target trials to estimate causal effect of precision medicine (PM) algorithm versus different controls}. Patients are first screened according to their eligibility for the algorithm: based on their genomic characteristics patients are recommended a specific treatment (eligible) or not (no eligible). Then eligible patients are randomized and assigned either to PM-directed arm or to one of the alternative control arms ($\text{CE}_1$, $\text{CE}_2$ or $\text{CE}_3$)}
\label{fig1_Target_Trials} % \label works only AFTER \caption within figure environment
\end{figure}

On the basis of these clinical examples, we propose three different target trials and their corresponding causal estimates, the clinical relevance of which may vary according to medical contexts. Each target trial contains a precision-medicine directed arm in which patients are treated in accordance with the precision medicine algorithm recommendations but they are differentiated from each other by alternative control arms (Figure~\ref{fig1_Target_Trials}). Causal effects will be estimated solely on patients eligible for the assignment of a personalized treatment, i.e. those for whom the treatment algorithm is able to recommend a drug. 

\subsection{First causal effect ($\textrm{CE}_1$): comparison with a single standard}\label{C1}

The first possible target trial is to compare the precision medicine arm with a control arm in which all patients have been treated with the same single treatment. This could classically be the current standard of care applied to all patients (e.g chemotherapy cancer treatment). 

\subsection{Second causal effect ($\textrm{CE}_2$): comparison with physician's assignment of drugs}\label{C2}

Then, in order to propose a more comprehensive clinical assessment, we propose a second causal effect, comparing the PM arm with the current clinical practice, i.e the assignment of the same targeted treatments by physicians in the absence of the algorithm. This implicitly means comparing two PM strategies: the one derived from the algorithm and the one that corresponds to current physician's knowledge. Unlike the former, the latter may not be perfectly deterministic depending on the heterogeneity of medical knowledge or practices. This way of defining $\textrm{CE}_2$ by focusing on the doctor's assignment of the same treatments stems from our question of interest: to quantify the relevance of the algorithm itself. Another vision would have been to compare the precision medicine arm with the doctor's treatments, allowing him to use treatments other than those of the PM arm, such as the gold-standard one described in $\textrm{CE}_1$. But the differences between the arms could then be biased by the use of treatments with different overall efficacy, changing the focus of the question. We will therefore stick to the first definition, which is more focused on the relevance of the algorithm. 

\subsection{Third causal effect ($\textrm{CE}_3$): comparison with random assignment of drugs}\label{C3}

Finally, we define the $\textrm{CE}_3$ effect comparing the PM arm with a control arm using exactly the same pool of treatments assigned randomly. In this case, we measure the ability of the PM algorithm to assign treatments effectively based on genomic features of patients. This comparison has already been considered in the context of biomarker-based clinical trials \cite{sargent2005clinical}. Although this comparison with random assignment is methodologically relevant, it may not make sense from a clinical point of view if the common clinical practice already contains  strong indications (or contraindications) for some patient-treatment associations.

\section{Precision medicine causal inference with multiple versions of treatment}

\subsection{Causal inference and the potential outcomes framework}\label{causal_simple}

Now that we have defined what we want to estimate, we need to specify the method of estimation. In cases where it would be too difficult or too early to conduct a true clinical trial we propose methods using observational data to emulate them. Indeed, it is possible to derive estimates with a causal interpretation from observational data in the context of the potential outcomes framework \cite{rubin1974estimating}. We will first describe briefly the fundamentals of this framework before extending this framework to our precision medicine settings in section \ref{causal_multiple}.

\subsubsection{First notations and causal graph\\}

\begin{figure}[ht]
\centering\includegraphics[width=0.6\textwidth]{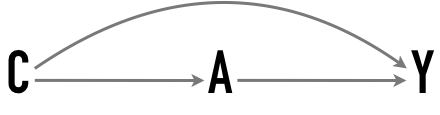}
\caption{\textbf{Causal diagram illustrating relations between variables}. Treatment A, outcome Y, and confounding variable C.}
\label{fig2_DAG_single} % \label works only AFTER \caption within figure environment
\end{figure}

We will use $j=1,...,N$ to index the individuals in the population. $A_j$ and $Y_j$ correspond respectively to the actual treatment received by individual $j$ and the outcome. In the most simple case, treatment takes values in $\mathcal{A}=\{0, 1\}$, $1$ denoting the treated patients and $0$ the control ones. $Y_j$ corresponds to the patient's response to treatment. In the case of cancer it may be  a continuous value (e.g size of tumour), a binary value (e.g status or event indicator), or even a time-to-event (e.g time to relapse or death). Only the first two cases will be discussed later. Finally, it is necessary to take into account the possible presence of confounders influencing both $A$ and $Y$ and denoted $C_j$ for individual $j$. These simple relations can be summarized in a causal graph (Figure~\ref{fig2_DAG_single}).

\subsubsection{Potential outcomes framework\\}

One standard framework to estimate causal effects relies on potential outcomes \cite{rubin1974estimating}. This framework is sometimes described as counterfactual because it defines variables like $Y_j(a)$ to denote the potential outcome of individual $j$ in case he has been treated by $A=a$ which may be different from what we observe if $A_j\neq a$. These counterfactual variables make it possible to write the causal estimands. For instance, in this context, we can easily compute the difference in outcome between treated patients and control patients: 
$E[Y | A=1] - E[Y | A=0].$

However, this difference has no causal interpretation as it does not offer any guarantees as to the confounding factor, as an unbalanced distribution of $C$ can induce biases Thus we define another estimate:
$E[Y(1)] - E[Y(0)].$

In this case, we compare between two ideal cohort, one in which all patients have been treated (possibly contrary to the fact) and one in which all patients have been left in the control arm (once again, possibly contrary to the fact). Under certain assumptions of consistency, positivity and conditional exchangeability, the potential outcomes framework allows to estimate these counterfactual variables and therefore infer causal estimates from observational (non-randomized) data \cite{rubin1974estimating, hernan2020causal}. 

\emph{Consistency} means that values of treatment under comparison represent well-defined interventions which themselves correspond to the treatments in the data:
$\textrm{if} \: A_j=a,  \textrm{then} \: Y_j(a)=Y_j.$

\emph{Exchangeability} means that treated and control patients are exchangeable, i.e if the treated patients had not been treated they would have had the same outcomes as the controls, and conversely. Since we usually observe some confounders we define conditional exchangeability to hold if cohorts are exchangeable for same values of confounding $C$. Therefore conditional exchangeability will hold if there are no unmeasured confounding:
$Y(a) \indep A | C.$

\emph{Positivity} assumption states that the probability of being administered a certain version of treatment conditional on $C$ is greater than zero:
$\textrm{if} \: P[C=c] \neq 0, P[A=a | C=c] >0.$
Intuitively, this positivity condition is required to ensure that the defined counterfactual variables make sense and do not represent something that cannot exist.

\subsubsection{Identification of causal effects\\}

Different methods provide estimators to evaluate causal effect from observational data. Throughout the article, we will describe essentially one method called standardization or parametric g-formula. Details on other types of estimators are available in Supplementary Materials, sections B and C. In this simple case (Figure~\ref{fig2_DAG_single}), the causal effect of treatment $A$ can be written with standardized means (formal proof in Supplementary Materials, section A):

\begin{equation}\label{standardized_equation}
E[Y(A=1)] - E[Y(A=0)] = \sum_{c} \Big(E[Y | A=1, C=c]-E[Y | A=0, C=c]\Big) \times P[C=c].
\end{equation}

Computationally, non-parametric estimation of $E[Y | A=a, C=c]$ is usually out of reach. Thus, on real-world dataset, $E[Y | A=a, C=c]$ is estimated through modelling and explicit computation $P[C=c]$ is replaced by its empirical estimate.

\subsection{Precision medicine and the multiple versions of treatment}\label{causal_multiple}

\subsubsection{A treatment with multiple versions\\}

The statement of the theoretical framework implicitly implies the uniqueness of the versions of the treatment \cite{rubin1980randomization} or at least the treatment variation irrelevance \cite{vanderweele2009concerning}. In the precision medicine case, the multiplicity of versions is inherent: a given treatment status may encompass several drugs since a patient may be associated with several molecular agents based on his or her genomic characteristics. $A$ can be seen as a compound treatment \cite{hernan2011compound} or a treatment with multiple versions \cite{vanderweele2013causal}.

Therefore, we define a variable $K_j$ denoting the version of treatment administered to individual $j$. If $A_j=a$ is the arm to which the patient is assigned, $K^a_j$ is the molecule received, the version of treatment $A=a$ (e.g a specific anti-cancer drug) and $K^a_j \in \mathcal{K}^a$, the set of versions of treatment $A=a$. In our precision medicine problem, $A=0$ will denote control patients and $A=1$ the patients treated with an anti-cancer drug of the precision medicine pool.  $\mathcal{K}^1=\{k^1_1, ..., k^1_P\}$ is the set of $P$ possible targeted treatments for $A=1$ patients. For the sake of simplicity we will assume that there is only one treatment version for $A=0$ controls, $\mathcal{K}^0=\{k^0\}$. %$\mathcal{K}^{1}_{PM}$ is the set of versions of treatment $A=1$ used in the PM strategy and $\mathcal{K}^{a}_{\text{CE}_i}$ the set of versions used in the control arm of $\text{CE}_i$. Therefore,  $\mathcal{K}^{0}_{CE_1}=k^0$, then $\mathcal{K}^{1}_{CE_2} \subseteq \mathcal{K}^1$, and $\mathcal{K}^{1}_{CE_3}=\mathcal{K}^{1}_{PM} \subseteq \mathcal{K}^1$.
%JB: I am not sure we are using the above notations. So I comment them and let's see

\begin{figure}[ht]
\centering\includegraphics[width=0.8\textwidth]{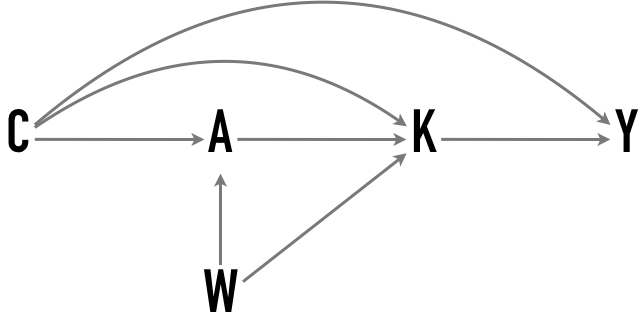}
\caption{\textbf{Causal diagram illustrating relations between variables under multiple versions of treatment}. Treatment A, version of treatment K, outcome Y, and confounding variables C and W.}
\label{fig3_DAG_multiple} % \label works only AFTER \caption within figure environment
\end{figure}

We also need to define other counterfactual variables like $K^a_j(a)$, the counterfactual version of treatment $A=a$ if the subject had been given the treatment level $a$. 
%Please note that in this precision medicine situation $K^1_j(1)=r(CE_j)$ in a purely deterministic way. 
Thus, we finally write the counterfactual outcome as $Y_j(a,k^a)$ for individual $j$ when treatment $A$ has been set to $a$, using $k^a$ as the version of treatment $a$, with $k^a \in \mathcal{K}^a$. Causal relations between variables $C$, $A$, $K$ and $Y$ are depicted in the causal diagram in Figure~\ref{fig3_DAG_multiple}. It should be noted that $A$ has no direct influence on $Y$, its only effect is entirely mediated by $K$, which is the real treatment in the pharmacological sense.

In this context, we can also define the assignment of a version of treatment for patients eligible to precision medicine algorithm. It is important to note that not all patients are necessarily eligible for the precision medicine strategy. Indeed, the treatment assignment algorithm relies on targetable alterations to establish its recommendations. In the absence of these, no recommendation can be offered to the patient. We denote $\mathcal{C}^{PM}$ the set of eligible patient profiles and consequently define the drug assignment algorithm as the function $r$ which associates to each $C$ a precision medicine treatment version $K$ such as:

$$\forall j \in \llbracket 1, N \rrbracket, \: \textrm{if} \: C_j \in \mathcal{C}^{PM}, r(C_j) \in \mathcal{K}^{1}$$

\subsubsection{Causal inference with multiple versions\\}

Consequently, the multiplicity of versions prevents direct application of the framework as described in section~\ref{causal_simple}. The theoretical framework has however been extended to causal inference under multiple versions of treatment and some identifiability conditions and properties have been studied, especially in the seminal article by \cite{vanderweele2013causal}. One of the first required adaptation to identify some causal effects is to partition confounders $C$ and $W$ (Figure~\ref{fig3_DAG_multiple}). $W$ indicates a collection of covariates that may be causes of treatment $A$ or version of treatment $K$ but are not direct causes of $Y$. These covariates are of special interest for causal effects identification under multiple versions of treatment. $C$ indicates all other covariates. In our precision medicine settings, the genomic features of patients may define the eligibility to precision medicine and therefore affect $A$. They may also be used to define the version of treatment $K$. And finally they can influence the response to treatment $Y$. Thus, the genomics features of patients, as defined in section \ref{introduction}, are a typical example of type $C$ confounders. All causal relationships are summarized in Figure~\ref{fig3_DAG_multiple}. Please note that all subsequent definitions are made taking into account W. However, no such variable is present in the application provided.

We summarize here some general observations from \cite{vanderweele2013causal} regarding the extension of the framework to multiple versions before discussing specific estimates of interest of our precision medicine settings in section~\ref{identification_causal_PM}. First of all, the identifiability conditions have to be adapted. The \emph{consistency} assumption for instance is extended to $K$:

$$\textrm{if} \: A_j=a,  \textrm{then} \: K_j^a(a)=K_j^a$$.

Then, the \emph{conditional exchangeability} or no-unmeasured confounding assumptions, may be stated in two different ways, either without or with versions of treatment:

\begin{equation}\label{exchan_noK}
Y(a) \indep A | (C,W)
\end{equation}

\begin{equation}\label{exchan_K}
Y(a, k^a) \indep \{A, K\} | C
\end{equation}

If equation \ref{exchan_noK} holds, we can derive a new version of the standardised estimator (Equation~\ref{standardized_equation}) with multiple versions of treatment \cite{vanderweele2013causal}:

\begin{equation}\label{overall_treatment_effect}
E[Y(a)] = E[Y(a, K^a(a))] = \sum_{c,w} E[Y | A=a, C=c, W=w] \times P[c,w]
\end{equation}

Specifically, it should be noted that we need to add $W$ in the set of covariates that must be taken into account in standardization, and we need \emph{positivity} to hold for $C$ and $W$, i.e. $0<P[A=a|C=c,W=w]<1$. Detailed proof of equation~\ref{overall_treatment_effect} is provided in Supplementary Materials, section A. Equation~\ref{overall_treatment_effect} paves the way to overall treatment effect assessment since $E[Y(1, K^1(1))]-E[Y(0, K^0(0))]$ would estimate the effect of treatment $A=1$ compared to $A=0$ with current versions of treatment.

Conversely, estimating a treatment effect for a given unique version of treatment $E[Y(a,k^a)]$ would require to control for exchangeability with versions K and therefore to hold equation~\ref{exchan_K} true \cite{vanderweele2013causal}:

\begin{equation}\label{version_treatment_effect}
E[Y(a, k^a)] = \sum_{c} E[Y | A=a,K^a=k^a, C=c] \times P[c]
\end{equation}

Similarly, we can define $G^a$ a random variable for versions of treatment with conditional distribution $P[G^a=k^a| C=c]=g^{k^a,c}$ and assuming the equation~\ref{exchan_K} to be true we can derive the following formula and its formal proof in Supplementary Materials, section A:

\begin{equation}\label{distrib_treatment_effect}
E[Y(a, G^a)] = \sum_{c,k^a} E[Y | A=a,K^a=k^a, C=c] \times g^{k^a,c} \times P[c]
\end{equation}

In this case, to allow estimation of the right-hand side of the equation, positivity will be defined as $0<P[A=a, K^a=k^a|C]<1$. 

\subsubsection{Application to precision medicine\\} \label{identification_causal_PM}

In the context of the potential outcomes framework extended to treatments with multiple versions, it is therefore possible to apply equations~\ref{overall_treatment_effect} and \ref{distrib_treatment_effect} in order to define and estimate the precision medicine causal effects previously described in section~\ref{causal_estimates}.

$A=0$ corresponds to control patients with $\mathcal{K}^0=\{k^0\}$ and $A=1$ to patients treated with a targeted treatments. It is important to notice that from this point on we systematically restrict ourselves to patients eligible to the precision medicine algorithm, i.e. to individuals $j$ such as $C_j \in \mathcal{C}^{PM}$. 

\paragraph{$\textrm{CE}_1$ estimation\\}

$\textrm{CE}_1$ is a comparison between the precision medicine arm and a single version control arm:

\begin{equation}
    \begin{aligned}
        \textrm{CE}_1 &= E[Y(1, r(C)] - E[Y(0, k^0)]
    \end{aligned}
\end{equation}

In details, $E[Y(1, r(C)]$ can be derived from equation~\ref{distrib_treatment_effect} in the case where $g^{k^a,~c}=1$ if $k^a=r(c)$ and $g^{k^a,~c}=0$ otherwise:

\begin{equation*}
        E[Y(1, r(C)] = \sum_{c} E[Y | A=1,K^1=r(c), C=c] \times P[c]
\end{equation*}

Then, $E[Y(0, k^0)]$ and $E[Y(1, k^1_{ref})]$ can be derived from equation~\ref{version_treatment_effect}:

\begin{equation*}
        E[Y(0, k^0)] = \sum_{c} E[Y | A=0, C=c] \times P[c]
\end{equation*}

Alternatively, if one wants to use as control only one of the treatments used in the PM arm  the previous estimate could be replaced by the following one:
\begin{equation*}
        E[Y(1, k^1_{ref})] = \sum_{c} E[Y | A=1,K^1=k^1_{ref}, C=c] \times P[c]
\end{equation*}

It should be noted that $\textrm{CE}_1$, like $\textrm{CE}_2$ and $\textrm{CE}_3$ presented later, depends on the PM algorithm of interest $r$. $\textrm{CE}_i$ could therefore also be written $\textrm{CE}_1(r)$. 

\paragraph{$\textrm{CE}_2$ estimation\\}

Then, $\textrm{CE}_2$ is written using $K^1(1)$ the PM targeted treatment that would have been assigned to the patient by the physician if the patient had been allocated in arm $A=1$ with PM targeted treatments:

\begin{equation}
    \begin{aligned}
        \textrm{CE}_2 &= E[Y(1, r(C)] - E[Y(1, K^1(1))]
    \end{aligned}
\end{equation}

$E[Y(1, K^1(1))]$ is derived from equation~\ref{overall_treatment_effect}:

\begin{equation*}
        E[Y(1, K^1(1))] = \sum_{c,w} E[Y | A=1, C=c, W=w] \times P[c,w]
\end{equation*}

\paragraph{$\textrm{CE}_3$ estimation\\}

Defining $G^1$ as the random distribution of versions of treatment $k^1 \in \mathcal{K}^{1}$, $\textrm{CE}_3$ expresses as:

\begin{equation}
    \begin{aligned}
        \textrm{CE}_3 &= E[Y(1, r(C)] - E[Y(1, G^1)] && \textrm{with } P[G^1=k^1_i \in \mathcal{K}^1]=\dfrac{1}{|\mathcal{K}^1_{PM}|},\\
    \end{aligned}
\end{equation}

$|.|$ denoting the cardinality of the set. In this formula, $E[Y(1, G^1)]$ can be derived from equation~\ref{distrib_treatment_effect}:

\begin{equation*}
        E[Y(1, G^1)] = \dfrac{1}{|\mathcal{K}^1_{PM}|} \times \sum_{c,k^1_i} E[Y | A=1, K^1=k^1_i, C=c] \times P[c]
\end{equation*}

\subsection{Alternative estimation methods}

For the sake of simplicity and brevity, we detailed the standardization. However, other popular candidate methods can be used. Estimators based on the inverse probability weighting (IPW) and targeted maximum likelihood estimation (TMLE) will also be computed in the following sections. A description of the theoretical framework of these two approaches and their adaptation to multiple versions of treatment is provided in Supplementary Materials, sections B and C.

\subsection{Code implementation}

Computation of causal effects is implemented in R and the code is provided in the form of R notebooks (simulated data and PDX data) as well as in the form of an RShiny application (simulated data only). All of these files are available in the dedicated GitHub: \url{https://github.com/JonasBeal/Causal_Precision_Medicine}.

\section{Simulation study}
\label{simulation}

The proposed methods are first tested on simulated data in order to check the performance of the estimators in finite sample sizes.

\subsection{General settings}

\begin{table}
\caption{Intercepts and and linear coefficients in the linear models specified to simulate data.}
\label{TableSim}
\begin{center}
\begin{tabular}{@{}cccc@{}}
\hline
\multirow{2}{*}{Response variable} & \multirow{2}{*}{Intercept} & \multicolumn{2}{c}{Linear regression coeff.} \\
&& $Y \sim C_1$ & $Y \sim C_2$ \\ 
\hline
$Y(0, k^0)$ & 0 & 0 & 15 \\
$Y(1, k^1_1)$ & -25 & -15 & 10 \\
$Y(1, k^1_2)$ & 0 & 0 & -20 \\
\hline
\end{tabular}
\end{center}
\end{table}

Using the R package \emph{lava}, we simulate a super-population of 10000 patients  with variables C, A, K and Y as in \ref{fig3_DAG_multiple}. We first define two independent binary variables $C_1$ and $C_2$, representing mutational status of genomic covariates, with a prevalence of 40\%. By analogy with the PDX data presented in the next section, Y mimics the evolution of tumour volume and a low value (\emph{a fortiori} negative) corresponds to a better response. Y is therefore defined as a continuous gaussian variable. For each counterfactual variable of response $Y(a, k^a)$, we specify the intercept and the linear regression coefficients regarding influence of $C_i$ as described in Table~\ref{TableSim}. Lower intercepts correspond to better responses/more efficient drugs. Similarly, a negative regression coefficient between $Y(a, k^a_i)$ and $C_j$ means that the gene $C_j$ improves the response to $k^a_i$. So all in all, $k^1_1$ has the best basal response (lowest intercept). $C_1$ (resp. $C_2$) improves the response to $k^1_1$ (resp. $k^1_2$). The treatment algorithm of precision medicine is in line with these settings since patients mutated in $C_1$ (regardless their $C_2$ status) are recommended to take $k^1_1$ and patients mutated for $C_2$ only are recommended to take $k^1_2$. Patients without mutations are not eligible to precision medicine and not taken into account in the computations. Since $k^1_1$ has the bast basal response we assume it is assigned with greater probability by the physician and implement the following distribution of observed treatments:

$$P[K=k^1_1]=0.5 \textrm{ and } P[K=k^1_2]=P[K=k^0]=0.25$$

A super-population of 10000 patients is then generated. 1000 cohorts of 200 patients are sampled without replacement within this super-population which, with the prevalences defined for the mutations, corresponds to an effective sample size of about 130 patients eligible to the PM algortithm. the causal effects $\textrm{CE}_1$, $\textrm{CE}_2$ and $\textrm{CE}_3$ are computed based on different methods on the sub-cohort eligible to precision medicine:
\begin{itemize}
  \item True effects, using counterfactuals for all patients
  \item Naive effect, using observed outcomes only for both arms
  \item Corrected effects: using observed outcome, computed with standardized estimators (Std), inverse probability weighting (IPW) and targeted maximum likelihood estimators (TMLE). Details about estimators of the last methods are available in Supplementary Materials, sections B and C.
\end{itemize}

\subsection{Simulation results}

\begin{figure}[ht]
\centering\includegraphics[width=\textwidth]{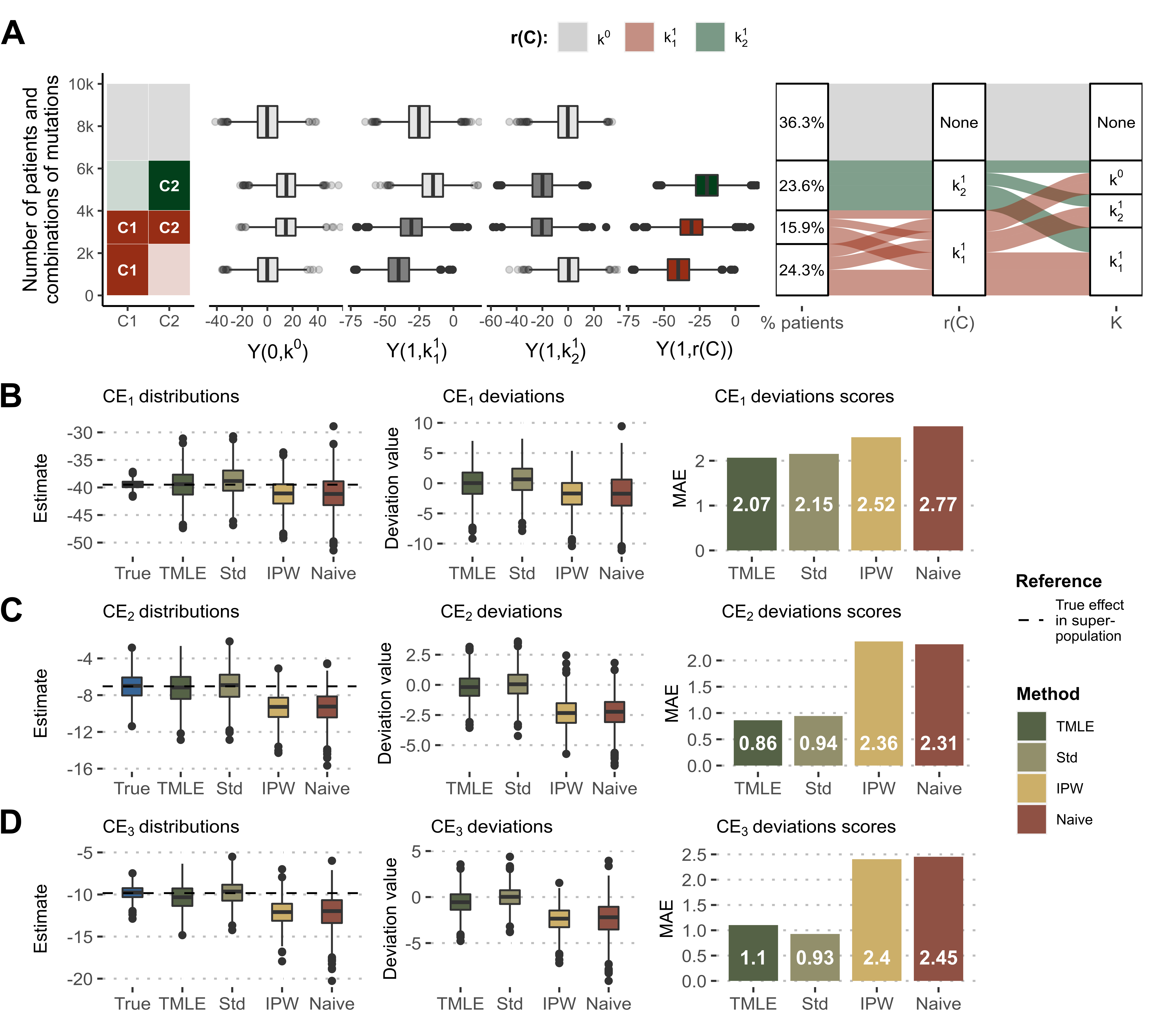}
\caption{\textbf{Causal effects of PM with simulated data}. (A) Main variables and relations in the simulated super-population. From left to right: categories of patient based on their mutations; responses to $k^0$, $k^1_1$, $k^1_2$ and precision medicine $K=r(C)$; repartition of patients regarding their precision medicine drug and their assigned treatment in observed data. (B) Distribution and deviation of $\text{CE}_1$ estimates based on different methods, deviation scores being computed based on mean absolute error (MAE). (C) Same for $\text{CE}_2$. (D) Same for $\text{CE}_3$. }
\label{Simulation_Results} % \label works only AFTER \caption within figure environment
\end{figure}

First, the distribution of data in the super-population of 10,000 patients can be observed in Figure~\ref{Simulation_Results}A, illustrating the different relations and differences described above. In particular, $Y(1, k^1_1)$ (resp. $Y(1, k^1_2)$) is lower for $C_1$-mutated (resp. $C_2$-mutated) patients. It can also be seen that the response to precision medicine ($Y(1, r(C))$) differs according to the groups: patients mutated for $C_1$ only have the best response, followed by patients mutated for both $C_1$ and $C_2$ and patients mutated for $C_2$ only. There is therefore a heterogeneity of responses to PM which encourages to take into account the groups of patients and their PM versions. The right side of Figure~\ref{Simulation_Results}A shows the deterministic assignment of the recommended PM treatment ($r(C)$) to each patient profile and the unbalanced distribution of observed treatments ($K$) with a predominance of $k^1_1$.

In the first target trial, true $\textrm{CE}_1$ estimates in the sampled cohorts are distributed around -40 (Figure~\ref{Simulation_Results}B), confirming the superiority of the PM arm over the control arm as defined in the simulation parameters. Not all methods of estimating the causal effect perform equally well. The so-called naive estimate and the one based on IPW show a net bias. The over-representation of the most advantaged patients by PM tends to cause these methods to overestimate the benefit of PM, as can also be seen in the deviation plots. The same trends are observed for $\textrm{CE}_2$ and $\textrm{CE}_3$ (Figure~\ref{Simulation_Results}C and D) where the differences are even more drastic. The mean absolute error of the naive method is thus divided by more than 2 when using standardized estimates or the TMLE.

In order to further dissect the influence of simulation parameters on estimation performances, a slightly different simulation scenario with equal probabilities of observed treatments is provided in Supplementary Materials, section D. This case shows that a random and balanced assignment of the observed treatments logically removes the systematic biases of the naive method by providing them with more randomized data. However, the corrections made by the proposed methods of causal inference, and in particular standardization and TMLE, reduce the variances in the estimates due to the heterogeneity of the effects of precision medicine as a function of molecular profiles.

\section{Application to pre-clinical data: patient-derived xenografts (PDX)}

The method is then applied to public data from patient-derived xenografts \cite{hidalgo2014patient}. Each patient tumour is divided into pieces later implanted in several immunodeficient cloned mice treated with different drugs, thus providing access to sensitivities to several different drugs for each tumour (see Supplementary Materials, section E). The original dataset contains 281 different tumours of origin (sometimes called PDX models, in the sense of a biological model) and 63 tested drugs, not all drugs having been tested for all tumours \cite{gao2015high}. 192 of these tumours have also been characterized for their mutations, copy-number alterations and mRNA.

Such data provides access to treatment response values  otherwise considered as hypothetical (or counterfactual). Availability of these data provides a unique ground truth to assess the validity of proposed causal estimates in a pre-clinical context. Based on the analysis accompanying the published data \cite{gao2015high}, some biomarkers of treatment response have been selected and resulted in an example of a treatment algorithm: binimetinib (MEK inhibitor) is recommended to KRAS/BRAF mutated tumours, and BYL719 (alpha-specific PI3K inhibitor, also known as Alpelisib) to PIK3CA mutated tumours. PTEN is also included as a covariate because of its detrimental impact on the response to these two treatments. LEE011 drug (a cell cycle inhibitor also known as Ribociclib) is chosen as the reference drug treatment ($k^0$). It should be noted that different drug response metrics are computed in the source data, two of which will be used in this case study. The first one is continuous and called \emph{BestAverageResponse} in the data, it is based on the variation of the tumour volume after treatment, the lower values (and especially negative) corresponding to better responses. The second one is originally categorical and based on a modified Response Evaluation Criteria In Solid Tumors (RECIST) criteria. It was binarized for this study so that the responders have a score of 1 and non-responders 0. The details of the definition and distribution of these metrics are given in Supplementary Materials, section E. Among the sequenced tumours, 88 are eligible to this precision medicine algorithm (\emph{i.e.} mutated for BRAF, KRAS or PIK3CA) and have been tested for all 3 drugs of interest, thus ensuring a the availability of all responses of interest. The following analyses will focus exclusively on this sub-cohort for which a comprehensive analysis is provided in Supplementary Materials, section E.

The analysis settings are similar to the ones used for simulated data. 1000 different cohorts of 70 tumours (out of 88) are sampled without replacement assuming each time that only the response to one of the treatments is known for each tumour, reproducing the classical clinical situation. The distribution of the observed treatments was defined randomly: $$P[K=k^0]=P[K=k^1_1]=P[K=k^1_2]=\dfrac{1}{3}$$
It should be noted that, contrary to analyses based on simulated data, all the statistical models used for standardisation (outcome model), for the IPW (treatment model) and for the TMLE are no longer generalized linear models (GLM) but random forests. This is intended to avoid misspecification due to the largely non-linear nature of the data. It was also observed that the performance of GLM-based methods was indeed lower than that of the naive method, supporting the importance of relevant model specification consistent with real data.

\begin{figure}[ht]
\centering\includegraphics[width=\textwidth]{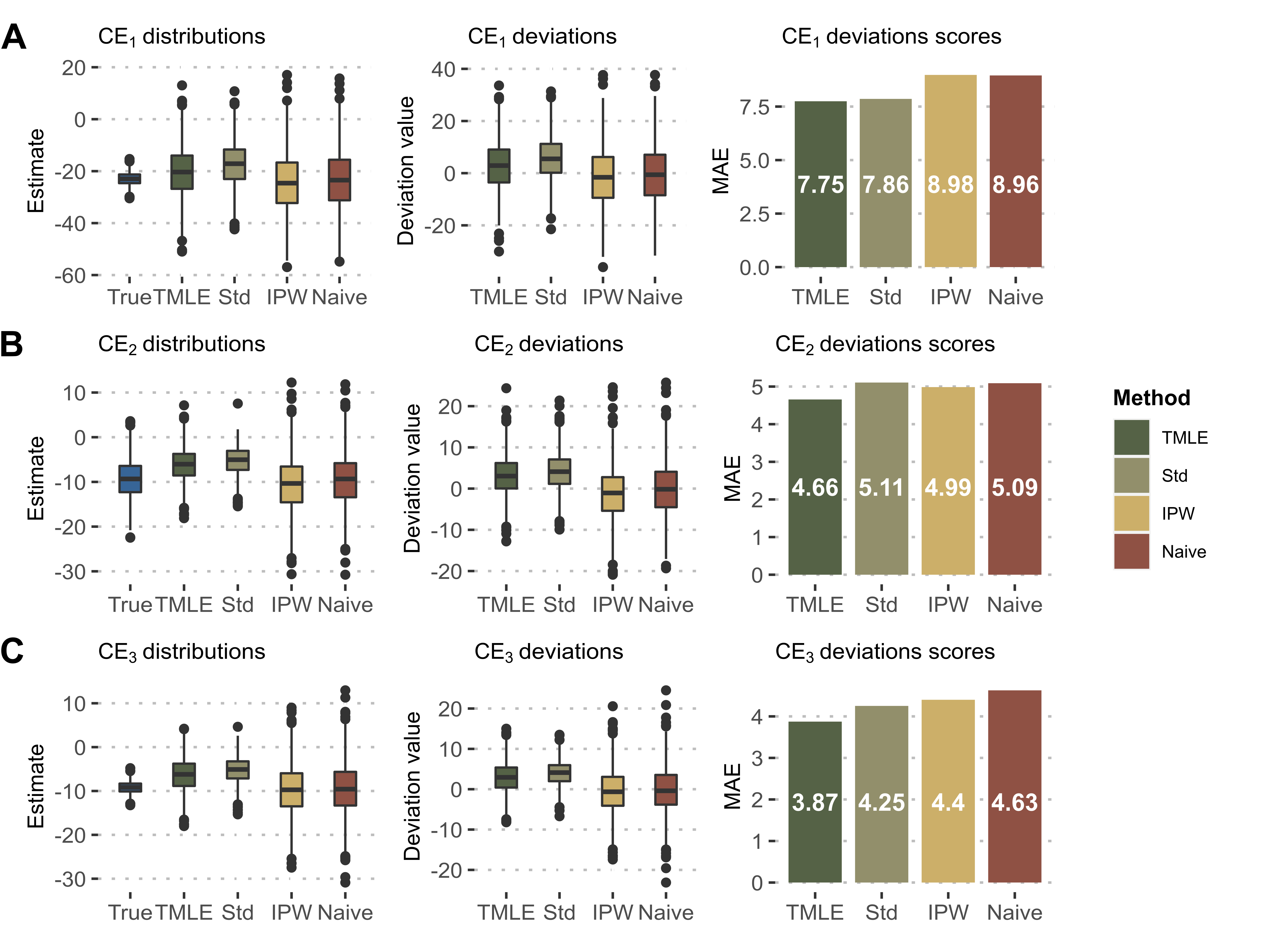}
\caption{\textbf{Causal estimates with PDX data}. Distribution and deviation of $\text{CE}_1$ (A), $\text{CE}_2$ (B) and $\text{CE}_3$ (C) estimates based on different methods}
\label{PDX_cont} % \label works only AFTER \caption within figure environment
\end{figure}

The results of estimations are then presented in Figure~\ref{PDX_cont}. In the presence of randomly assigned and balanced observed treatments, none of the methods (including the naive one) has significant systematic bias. On the other hand, more sophisticated methods, and in particular TMLE, allow to reduce the gap between estimates and true values, which is achieved through the special design of the PDX data.

\section{Discussion}

In synthesis, this work proposes a conceptual framework for evaluating a precision medicine algorithm, taking advantage of data already generated using adapted causal inference tools. In a clinical context, these data were not generated in a purely observational manner. Patients were cared for and treated by physicians who probably took into account some of their characteristics. However, the reasoning, formalized or not, behind the physicians' decisions does not correspond to that which a new investigator might want to test. In the eyes of this new investigator, the data can therefore be considered as observational in that they do not correspond to the randomization he would have liked to have carried out. The possibility for this new investigator to estimate the impact of his PM algorithm using the proposed estimators depends, however, on the consistency, exchangeability and positivity hypotheses.

The hypothesis of consistency has been made more plausible by taking into account the treatment versions, which makes it possible to explicit the heterogeneity of the molecules administered. Exchangeability remains questionable. The simulations and calculations described above underline the importance of taking into account at least the genomic covariates used in the processing algorithm. The inclusion of additional covariates is likely to be necessary in many real-world applications. Positivity, on the other hand, can be violated in a much more obvious way in certain situations. Thus, equation~\ref{distrib_treatment_effect} requires positivity to be extended to versions of treatment: $0<P[A=a, K^a=k^a|C]<1$. If the assignment of the observed treatments was done on a deterministic basis with respect to the variables used by the treatment algorithm, each patient's molecular profile will have been treated with a single drug, thus preventing any subsequent causal inference within the defined framework. The eventual use, by the boards of physicians in charge of assigning the observed treatments, of variables different from those used by the algorithm could then make it possible to verify the positive condition. But these variables would represent unmeasured confounding factors. It is therefore essential to have an in-depth knowledge of the rationales at work in the assignment of the observed treatments.

We developed a user-friendly application that extends  the scope of the simulations and makes possible to study and quantify the impact of different situations, including possible (quasi-)~violations of positivity or unmeasured confounding. It is thus a tool for empirically framing cases where this causal inference is reasonable or not. The analysis of the PDX data provides an illustration and proof of feasibility for these methods on pre-clinical data, closer to the human clinical data generally of interest. Beyond feasibility, this implementation leads to some remarks. Firstly, the improvement of causal inference methods compared to naive estimation of PM effects is conditioned in this case to the use of flexible and non-linear learning algorithms. This underlines the importance of a proper specification of the outcome and treatment models whose imperfection, especially when trained on small samples, could explain the modesty of the results compared to the simulated data. The particular nature of the PDX data design used should also be kept in mind: each tumour is tested only once for each drug, which may lead to greater variability of results due to tumour heterogeneity \cite{gao2015high}. Some studies, with smaller numbers of tumours and treatments, propose to form groups of several mice for each treatment-drug combination \cite{hidalgo2014patient}. The use of these mean effects could contribute to more accurate data. In spite of these limitations, which may diminish their ability to provide values with counterfactual interpretation, PDX data are thus a dataset of interest for studying and validating methods of causal inferences about treatment response. It can also be noted that the very nature of these data, due to the multiplicity of drugs tested for each tumour, can provide a framework in which the constraints of positivity are singularly alleviated. Even if all drugs were not tested on all patients, considering each tumour-drug combination as a different unit increases the coverage of the data. It is then necessary to take into account the clustered nature of the data, each tumour being present several times.

Finally, beyond the preclinical data presented here, the theoretical framework developed in this article should be more directly applicable to data from clinical trials if these data do not violate the requirements of positivity. If it is necessary to consider several trials, the heterogeneity of practices must be taken into account. The use of different drug lists from one trial to another or from one medical centre to another could also provide an example of confounding factor $W$, included in the theoretical framework presented here but not used in applications.

\section{Software}

All analyses have been performed with R. Analysis on simulated data is available as R code organized in a notebook and also as a RShiny interactive application designed to test different simulation scenarios. Analysis on PDX data is available as R notebook and the original dataset is also provided. All files and documentation are available on the dedicated GitHub repository: \url{https://github.com/JonasBeal/Causal_Precision_Medicine}. For the interactive application, readers unfamiliar with R can refer to the online version of the RShiny application: \url{https://jonasbeal.shinyapps.io/application_causal_pm/}.

\section{Supplementary Materials}

Theoretical details are given about standardised estimators in section A, about IPW estimators in section B and about TMLE in section C. Section D showcases an additional scenario with simulated data. An extensive analysis of PDX data is also provided in section E. Section F is focused on an example with a binary $Y$ outcome.

\section*{Acknowledgments}

The authors thank Christophe Le Tourneau, Stephen Cole and Margarita Moreno-Betancur for helpful comments and discussions.

{\it Conflict of Interest}: None declared.

\bibliographystyle{unsrt}  
%\bibliography{references}

\end{document}

\maketitle

\renewcommand{\thesection}{\Alph{section}}

\section{Formal proofs for standardized causal estimates}

The main notations and definitions have been described in the body of the article. Details of some equations are provided below.

\subsection{Unique version of treatment}

Here is the formal proof for equation 3.1:

\begingroup
\footnotesize
\begin{equation*}
\begin{aligned}
  E[Y(a)] & = \sum_{c} E[Y(a)|c] \times P[c]&& \\
          & = \sum_{c} E[Y(a)|a,c] \times P[c,w]
          &&\text{by conditional exchangeability} Y(a) \indep A | C \\
          & = \sum_{c} E[Y|a,c] \times P[c]
          &&\text{by consistency for Y}
\end{aligned}
\end{equation*}
\endgroup

\subsection{Overall treatment effect with multiple versions of treatment}

Here is the formal proof for equation 3.4, mostly derived from the proof of Proposition 3 in \cite{vanderweele2013causal}.

\begingroup
\footnotesize
\begin{equation*}\label{overall_treatment_effect_proof}
\begin{aligned}
  E[Y(a, K^a(a))] & = E[Y(a)] &&K^a \text{is the version actually received}\\
                  & = \sum_{c,w} E[Y(a)|c,w] \times P[c,w]&& \\
                  & = \sum_{c,w} E[Y(a)|a,c,w] \times P[c,w]
                  &&\text{with } Y(a) \indep A | (C,W)\\
                  & = \sum_{c,w} E[Y(a, K^a(a))|a,c,w] \times P[c,w]&& \\
                  & = \sum_{c,w,k^a} E[Y(a, k^a)|a,K^a(a)=k^a,c,w]\times P[K^a(a)=k^a|a,c,w] \times P[c,w] &&\\
                  & = \sum_{c,w,k^a} E[Y(a, k^a)|a,K^a=k^a,c,w]\times P[K^a=k^a|a,c,w] \times P[c,w]
                  &&\text{(consistency for K)} \\
                  & = \sum_{c,w,k^a} E[Y|a,K^a=k^a,c,w]\times P[K^a=k^a|a,c,w] \times P[c,w]
                  &&\text{by consistency for Y} \\
                  & = \sum_{c,w} E[Y|a,c,w]\times P[c,w]
\end{aligned}
\end{equation*}
\endgroup

Then, the overall treatment effect can be defined and computed by:

$$E[Y(a, K^a(a))] - E[Y(a^*, K^{a^*}(a^*))]$$

\subsection{Treatment effect with predefined distributions of versions of treatment}

Here is the formal proof for equation 3.6, partially derived from the proof of Proposition 5 in \cite{vanderweele2013causal}.

\begingroup
\footnotesize
\begin{equation*}\label{distrib_treatment_effect_proof}
\begin{aligned}
  E[Y(a, G^a)] & = \sum_{c} E[Y(a, G^a)|C=c] \times P[c]\\
             & = \sum_{c, k^a} E[Y(a, k^a)|G^a=k^a, C=c] \times P[G^a=k^a|C=c] \times P[c]\\
             & = \sum_{c, k^a} E[Y(a, k^a)| C=c] \times g^{k^a,c} \times P[c]
             &&\text{since } P[G^a=k^a] = g^{k^a,c}\\
             & = \sum_{c, k^a} E[Y(a, k^a)| A=a, K^a=k^a, C=c] \times g^{k^a,c} \times P[c] 
             &&\text{with } Y(a,k^a) \indep \{A,K\} | C\\
             & = \sum_{c, k^a} E[Y| A=a, K^a=k^a, C=c] \times P[c]
             &&\text{by consistency for Y}
\end{aligned}
\end{equation*}
\endgroup

\clearpage
\section{Theoretical framework for estimates based on inverse probability weighting (IPW)}

Based on the same counterfactual framework, it is possible to build another class of models, called marginal structural models \cite{robins2000marginal}, from which we derive estimators different from the standardized estimators called inverse-probability-of-treatment weighted (IPW) estimators \cite{cole2008constructing}.

\subsection{Generic notations and estimators}

IP weighting is equivalent to creating a pseudo-population where the link between covariates and treatment is cancelled. In the case of binary treatment $A \in {0 ,1}$, weights are defined for each patient as the inverse of the probability to have received the version of treatment he or she actually received, knowing his or her covariates:

\begin{equation*}
        W^A=\dfrac{1}{f[A|C]} \text{ with } f[a|c]=P[A=a|C=c],
\end{equation*}

$f[a|c]$ being called the propensity score, \emph{i.e.} the probability to have received the treatment $A=a$, given the covariates $C=c$. Under the same hypothesis of exchangeability, positivity and consistency we can derive the modified Horvitz-Thompson estimator \cite{hernan2020causal}:

\begin{equation*}
E[Y(a)]=\dfrac{\hat{E}[I(A=a)W^{A}Y]}{\hat{E}[I(A=a)W^A]},
\end{equation*}

$I$ being the indicator function. It is also possible to define stabilized weights that will be used in this work: 

\begin{equation*}
SW^A=\dfrac{f(A)}{f[A|C]}.
\end{equation*}

\subsection{Extension to multiple version}

An extension of these IPW methods to multi-valued treatments (only treatment $K$ with different modalities and no $A$) has already been studied and the different formulas and estimators adapted accordingly \cite{imbens2000role, feng2012generalized}, defining in particular a generalized propensity score:

\begin{equation*}
f(k|c)=P[K=k|C=c]=E[I(k)|C=c] 
\end{equation*}

\begin{equation*}
\text{with }
I(k) = \left\{
\begin{array}{ll}
1 & \quad \text{if } K = k \\
0 & \quad \text{otherwise}
\end{array}
\right.
\end{equation*}

and a subsequent estimator:

\begin{equation*}
E[Y(k)]=\dfrac{\hat{E}[I(K=k)W^{K}Y]}{\hat{E}[I(K=k)W^K]} \text{ with } W^K=\dfrac{1}{f[K|C]}
\end{equation*}

\subsection{Application to precision medicine estimates}

In our case, to be consistent with the previously defined causal diagram we have both $A$, binary status depending on the class of drugs, and $K$, the multinomial variable for versions of treatments, \emph{i.e.} the precise drug. Therefore we need to define a slightly different propensity score with joint probabilities:

\begin{equation*}
\begin{aligned}
f(a,k|c) & =P[A=a,K=k|C=c]
         & =P[K=k|A=a, C=c].P[A=a|C=c]
         & =E[I(a,k)|C=c] 
\end{aligned}
\end{equation*}

\begin{equation*}
\text{with }
I(a,k) = \left\{
\begin{array}{ll}
1 & \quad \text{if } A = a, K = k \\
0 & \quad \text{otherwise}
\end{array}
\right.
\end{equation*}

From this we can deduce the estimator:

\begin{equation*}
E[Y(a, k)]=\dfrac{\hat{E}[I(A=a,K=k)W^{A,K}Y]}{\hat{E}[I(A=a,K=k)W^{A,K}]} \text{ with } W^{A,K}=\dfrac{1}{f[A,K|C]}
\end{equation*}

In all the examples presented in this study and implemented in the code, $\mathcal{K}^{0} \cap \mathcal{K}^{1} = \emptyset$, it is therefore possible to simplify the joint probabilities since the knowledge of K automatically results in the knowledge of A allowing $P[A=a, K=k|C=c]=P[K=k|C=c]$. The above formulas with the attached probabilities are still necessary in the general case and allow for the derivation of causal effects $\text{CE}_1$, $\text{CE}_2$ and $\text{CE}_3$ previously described.
\clearpage

\section{Some elements about targeted maximum-likelihood estimation (TMLE)}

Targeted maximum likelihood estimation is framework based on a doubly robust maximum-likelihood–based approach that includes a "targeting" step that optimizes the bias-variance trade-off for a defined target parameter. In particular, this method is perfectly compatible with the use of machine learning algorithms for outcome or treatment models. A detailed description of the method and its implementations can be found in \cite{van2011targeted}.

The implementation proposed in this article is very similar to the one proposed in a recent tutorial concerning the application to binary processing \cite{luque2018targeted}. The specific characteristics of the problem of precision medicine studied here lead to modify this approach. In particular, the outcome and treatment models used in the first steps are modified in the same way as the one explained for the standardized estimators (outcome model) and for the IPW estimators (treatment model). The step of updating the estimates is done on a model similar to \cite{luque2018targeted}. 

The algorithm used for the models internal to the TMLE are, as much as possible, the same as those used for the standardised and IPW estimators. For simulated data generalized linear models have been used in all cases except multinomial classification andperformed through the function \emph{multinom} in \emph{nnet} package. For PDX data, random forests have been used for all models; use of SuperLearner is made possible by simple modifications to the code but significantly slows down its execution.  In the latter case random forests were chosen for their speed and versatility, especially in view of their ability to handle multinomial classification as well.

\clearpage

\section{An additional simulation scenario}

For this first scenario, observed treatments are assigned randomly with equal probabilities:
$$P[K=k^0]=P[K=k^1_1]=P[K=k^1_2]=\dfrac{1}{3}$$

\begin{figure}[!p]
\centering\includegraphics[width=\textwidth]{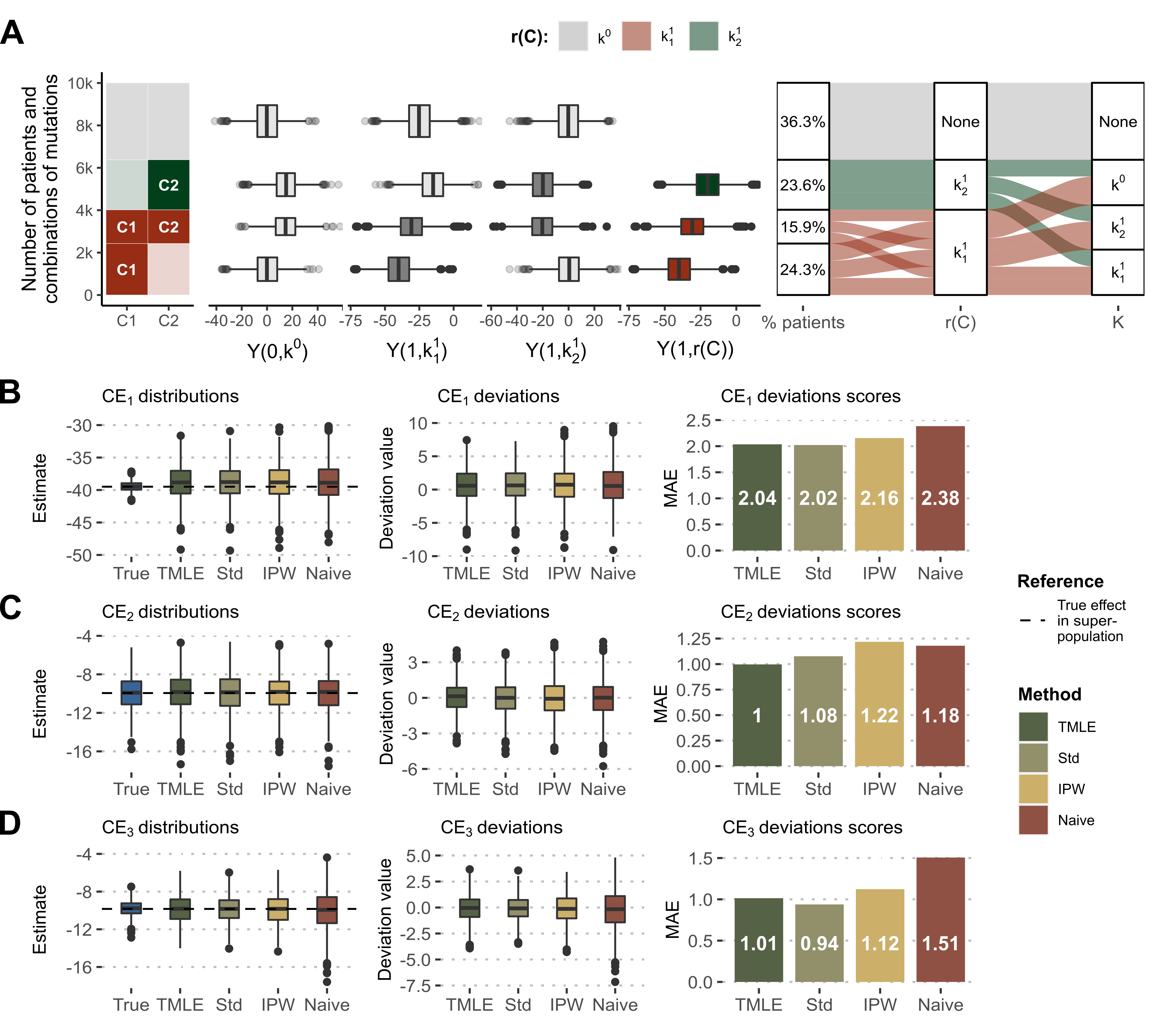}
\caption{\textbf{Causal effects of PM with simulated data}. (A) Main variables and relations in the simulated super-population. From left to right: categories of patient based on their mutations; responses to $k^0$, $k^1_1$, $k^1_2$ and precision medicine $K=r(C)$; repartition of patients regarding their precision medicine drug and their assigned treatment in observed data. (B) Distribution and deviation of $\text{CE}_1$ estimates based on different methods, deviation scores being computed based on mean absolute error (MAE). (C) Same for $\text{CE}_2$. (D) Same for $\text{CE}_3$. }
\label{Simulation_Additional} % \label works only AFTER \caption within figure environment
\end{figure}

Although this scenario is not necessarily the most clinically plausible, it allows us to observe the impact of treatment and response heterogeneity on effect estimates. The main features of generated data is described in Figure~\ref{Simulation_Additional}A. The results of estimation of causal effects using different methods are summarized in Figure~\ref{Simulation_Additional}B, C and D. First of all, we observe that the PM arm is estimated to be better than the 3 controls, which is logical with the way the data were simulated, with a treatment algorithm perfectly adapted to the data generation process. Besides, random assignment of the observed treatments prevents the naive method from being systematically biased, but it does not prevent it from demonstrating greater variance and mean absolute error.

\clearpage

\section{Additional information about patient-derived xenografts (PDX) data}

\begin{figure}[ht] %s state preferences regarding figure placement here
\centering\includegraphics[width=\textwidth]{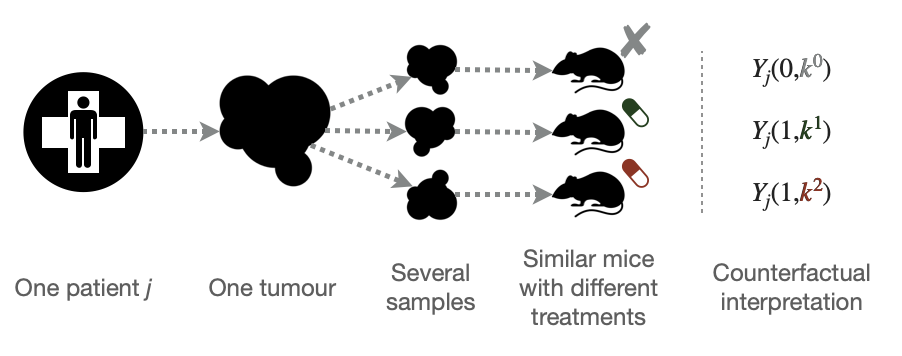}
\caption{\textbf{Principles of PDX screening}.  Schematic pipeline for PDX screening with tumour biopsies from one patient divides in several pieces later implanted in similar immunodeficient mice. Each mouse is then treated with a different drug. The collection of mice that have received tumour samples from the same patient but have been treated with different drugs therefore gives access to outcomes otherwise considered as Counterfactual. This pipeline follows the "one animal per model per treatment" approach ($1 \times 1 \times 1$) as described in \cite{gao2015high}.}
\label{PDX_principles} % \label works only AFTER \caption within figure environment
\end{figure}

Patient-derived xenografts have been used in different settings which nevertheless take up the same fundamentals: implanting a tumour from a patient in an immunodeficient mouse in order to observe its evolution and ideally its response to treatment. This therefore provides an intermediate pre-clinical model between the patient and the \emph{in vitro} cell lines that do not take into account the tumour micro-environment. A more precise description of these models and the different ways of using them can be found in dedicated reviews \cite{hidalgo2014patient} and the schematic representation that applies to the data presented here is described in Figure~\ref{PDX_principles}.

\subsection{PDX data and drug response metrics}

\subsubsection{A continuous outcome}

The first drug response metric used in this article is called \emph{Best Average Response}. For each combination tumour/drug, the response is determined by comparing tumor volume change at time $t$, $V_t$ to tumor volume at time $t_0$, $V_{t_0}$. Several scores are computed:

$$\text{Tumour Volume Change (\%)} = \Delta Vol_t = 100\% \times \dfrac{V_t-V_{t_0}}{V_t}$$

$$\text{Best Response} = min(\Delta Vol_t), t>10d$$

$$\text{Average Response}_t = mean(\Delta Vol_i, 0 \leq i\leq t)$$

$$\text{Best Average Response} = min(\text{Average Response}_t), t>10d$$

We will mainly focus on \emph{Best Average Response}. This metric "captures a combination of speed, strength and durability of response into a single value" \cite{gao2015high}. Qualitatively, lower values correspond to more efficient drugs.

\subsubsection{A binary outcome}

Thresholds of \emph{Best Response} and \emph{Best Average Response} are also defined, inspired by RECIST criteria \cite{therasse2000new}, in order to classify response to treatment into 4 categories: Complete Response (CR), Partial Response (PR, Stable Disease (SD) and Progressive Disease (PD). We designed a binary response status by combining the response categories (CR, PR and SD) into a single 'responder' category (1), opposed to the 'non-responders' progressive diseases (0).

\subsection{General description of the 88 PDX models sub-cohort}

Treatment assignment algorithm and observed drug sensitivities are consistent since mutated BRAF/KRAS tumours have a better response to binimetinib and mutated PIK3CA tumours have a better response to BYL719. In addition, it can be noted that these biomarkers have deleterious cross-effects.

\begin{figure}[ht] %s state preferences regarding figure placement here
\centering\includegraphics[width=\textwidth]{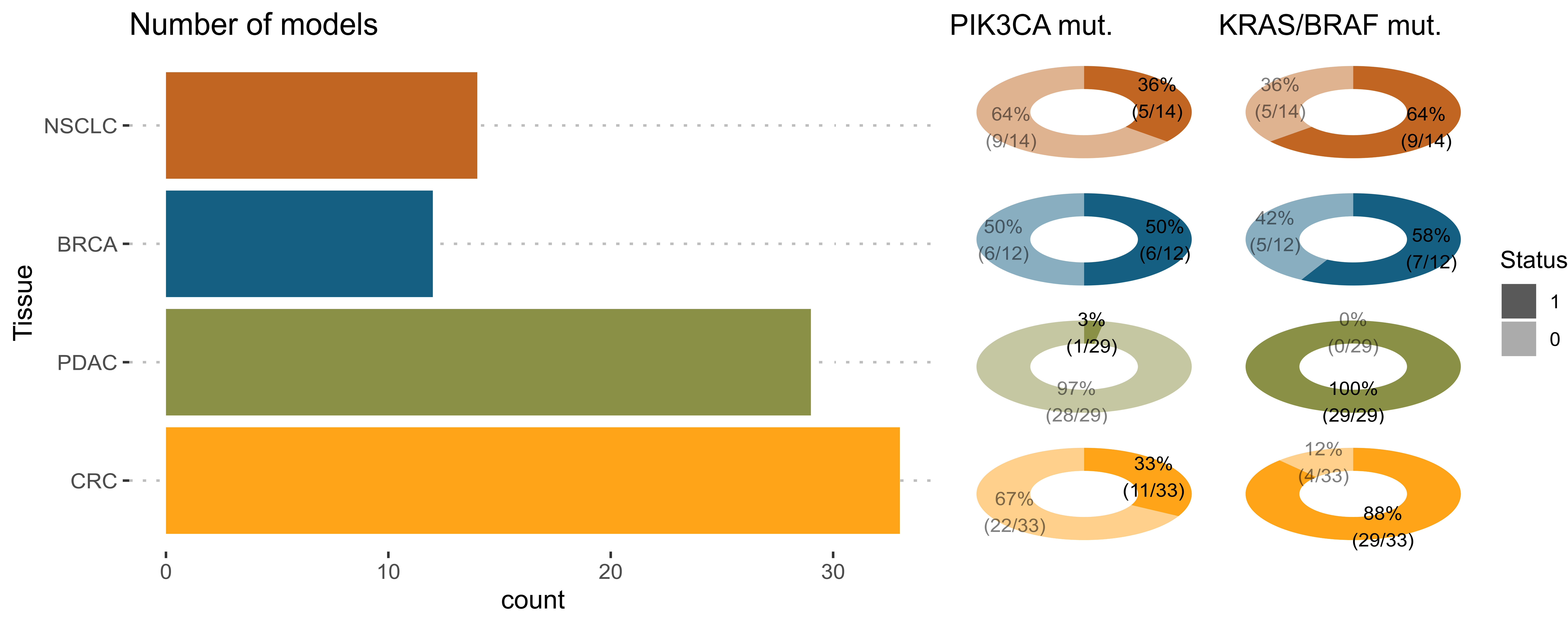}
\caption{\textbf{Description of the 88 PDX models cohort}. Tissue of origin and prevalences of the drug biomarkers.}
\label{PDX_distrib} % \label works only AFTER \caption within figure environment
\end{figure}

\begin{figure}[ht] %s state preferences regarding figure placement here
\centering\includegraphics[width=\textwidth]{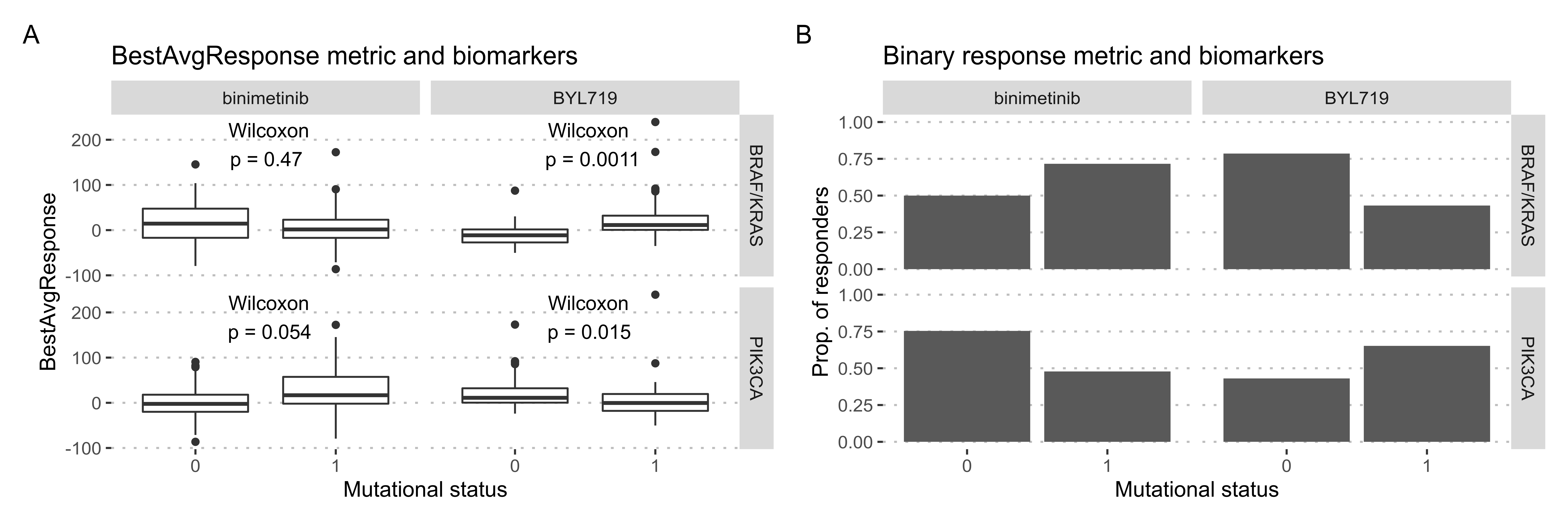}
\caption{\textbf{Drug response to PM treatments in the 88 PDX models cohort}. Continuous (A) and binary (B) response to binimetinib and BYL719 depending on the mutational status of biomarkers.}
\label{PDX_sensitivities} % \label works only AFTER \caption within figure environment
\end{figure}

\clearpage

\section{Causal estimates with a binary outcome}

An analysis similar to the one presented in section 6 of the paper is detailed here, replacing the continuous outcome with a binary response outcome. The definition of this binary outcome is given in the previous section.

\begin{figure}[ht]
\centering\includegraphics[width=\textwidth]{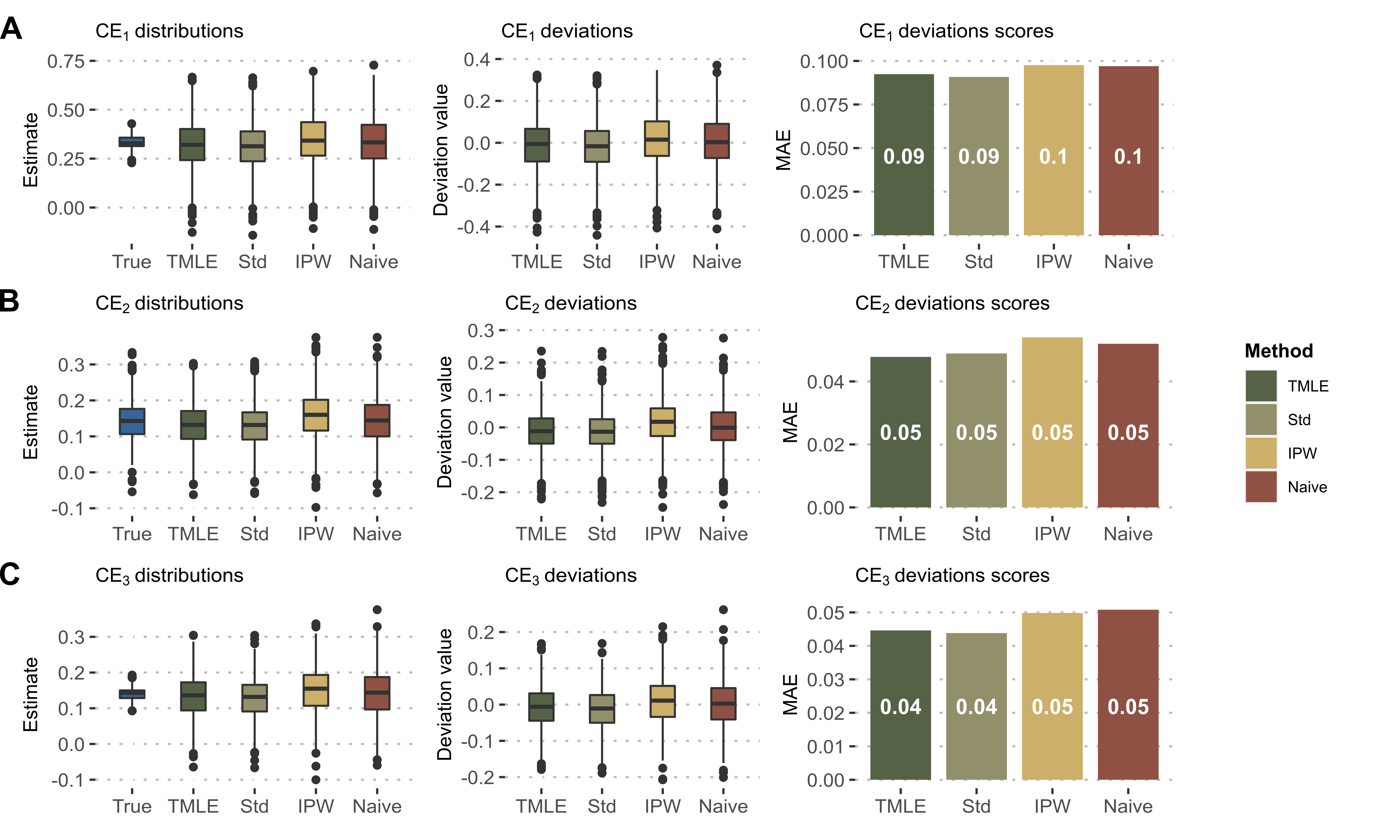}
\caption{\textbf{Causal estimates with PDX data binary outcome}. Distribution and deviation of $\text{CE}_1$ (A), $\text{CE}_2$ (B) and $\text{CE}_3$ (C) estimates based on different methods}
\label{PDX_bin} % \label works only AFTER \caption within figure environment
\end{figure}

\clearpage

\bibliographystyle{unsrt}
%\bibliography{references}